# Ultrahigh Pressure Superconductivity in Molybdenum Disulfide

Zhenhua Chi[1,2]*[†], Feihsiang Yen[1,2][†], Feng Peng[3][†], Jinlong Zhu[4][†], Yijin Zhang[5], Xuliang Chen[1,2], Zhaorong Yang[1,2,6], Xiaodi Liu[1], Yanming Ma[7], Yusheng Zhao[4], Tomoko Kagayama[8], Yoshihiro Iwasa[5,9]

[1]Key Laboratory of Materials Physics, Institute of Solid State Physics, Hefei Institutes of Physical Science, Chinese Academy of Sciences, Hefei 230031, China

[2]High Magnetic Field Laboratory, Hefei Institutes of Physical Science, Chinese Academy of Sciences, Hefei 230031, China

[3]College of Physics and Electronic Information, Luoyang Normal University, Luoyang 471022, China

[4]High Pressure Science and Engineering Center (HiPSEC) and Department of Physics & Astronomy, University of Nevada Las Vegas, Las Vegas, Nevada 89154, USA

[5]Quantum-Phase Electronics Center (QPEC) and Department of Applied Physics, University of Tokyo, Tokyo 113-8656, Japan

[6]Collaborative Innovation Center of Advanced Microstructures, Nanjing University, Nanjing 210093, China

[7]State Key Laboratory of Superhard Materials, Jilin University, Changchun 130012, China

[8]KYOKUGEN, Center for Quantum Science and Technology under Extreme Conditions, Graduate School of Engineering Science, Osaka University, Osaka 560-8531, Japan

[9]RIKEN Center for Emergent Matter Science (CEMS), Wako 351-0198, Japan

*Corresponding author. E-mail: zhchi@issp.ac.cn

[†]These authors contributed equally to this work.

**Superconductivity commonly appears under pressure in charge density wave (CDW)-bearing transition metal dichalcogenides (TMDs)[1,2], but has emerged so far only via either intercalation with electron donors[3] or electrostatic doping[4] in CDW-free TMDs. Theoretical calculations have predicted that the latter should be metallized through bandgap closure under pressure[5,6], but superconductivity remained elusive in pristine 2$H$-MoS$_2$ upon substantial compression, where a pressure of up to 60 GPa only evidenced the metallic state[7,8]. Here we report the emergence of superconductivity in pristine 2$H$-MoS$_2$ at 90 GPa. The maximum onset transition temperature $T_c^{(onset)}$ of 11.5 K, the highest value among TMDs and nearly constant from 120 up to 200 GPa, is well above that obtained by chemical doping[3] but comparable to that obtained by electrostatic doping[4]. $T_c^{(onset)}$ is more than an order of magnitude larger than present theoretical expectations, raising questions on either the current calculation methodologies or the mechanism of the pressure-induced pairing state. Our findings strongly suggest further experimental and theoretical efforts directed toward the study of the pressure-induced superconductivity in all CDW-free TMDs.**

The layered atomic structure of 2$H$-MoS$_2$ consists of a triangular plane of Mo atoms sandwiched between two triangular planes of S atoms in a trigonal prismatic coordination. The intra-layer and inter-layer interactions are covalent and van der Waals, respectively. Under high pressure, this material is well known to undergo a polymorphic structural transition [8,9] from 2$H_c$ to 2$H_a$, accompanied by bandgap closing suggestive of a semiconductor to metal transition [7,8] near 20–30 GPa. These findings agree very well with density functional calculations [5,6] which predicted a semimetallic phase following the bandgap closing but also a relatively poor superconducting propensity, at least of the BCS type as described by an averaged Eliashberg electron-phonon coupling parameter $\lambda$, at higher pressure [5]. We present here, for the first time, unambiguous experimental evidences of superconductivity in pristine MoS$_2$ at ultrahigh pressures, the insights into which call for a stringent analysis of more sophisticated possibilities.

The $R$-$T$ curves at pressures below 60 GPa and above 90 GPa are shown in Fig. 1a and Fig. 1b, respectively. Below 25 GPa, the behavior of the electrical resistance obeys the temperature dependence of a semiconductor. Above 25 GPa, a metallic state indicated by a positive d$R$/d$T$ slope between around 100 K and 200 K starts to develop within the parent semiconducting state which has a negative d$R$/d$T$ slope. The semiconducting state dominates below 60 GPa whereas the metallic state prevails between 60 GPa and 90 GPa. Above 90 GPa, the entire system completely transforms into a metallic state, out of which superconductivity, indicated by a precipitous drop in resistance, sets in at low temperature. The upturn in resistance which precedes the superconducting transition over the 90–150 GPa pressure range suggests either the presence of preformed pairs or localization of charge carriers, as observed in pressurized 1$T$-TiSe$_2$ (ref. 1) and 1$T$-TaS$_2$ (ref. 2). The disappearance of this upturn above 150 GPa might corresponds to delocalization of charge carriers. Zero resistance was inaccessible in our measurements owing to the contribution of contact resistance between the sample and the electrodes in the quasi-four-probe method. The onset transition temperature $T_c^{(onset)}$ is determined from the intersection of two extrapolated lines drawn through the resistance curve in the normal state just above the drop in resistance and through the steepest part of the resistance curve in the superconducting state. $T_c^{(onset)}$ is significantly enhanced from 5 K at 90 GPa to 9 K at 110 GPa and eventually saturates around 11 K over the pressure range of 120 GPa up to 200 GPa, much higher than the chemical doping case[3] but comparable to the electrostatic doping case[4].

To further substantiate the observed superconductivity, the effect of external magnetic field on the temperature dependency of the resistance of sample was investigated at 200 GPa under applied magnetic field up to 6.5 T, as shown in Fig. 2. The drop in the resistance is significantly suppressed with the increasing of the magnetic field until smearing out at 4.5 K and 5 T, indicating the type-II superconducting nature of the transition. The temperature-dependent upper critical magnetic field $\mu_0 H_{c2}$ (defined at 90％ of the normal state resistance at the transition temperature $T_c$) is shown in the inset. The $\mu_0 H_{c2}$ extrapolated to zero temperature using conventional one-band Werthamer-Helfand-Hohenberg (WHH) formula[10] gives $\mu_0H_{c2}(0)$=6.29 T, smaller than that of 7.33 T extrapolated by Ginzburg-Landau equation, larger than that of 4 T for $Bi_2Se_3$ at 30 GPa (ref. 13).

The *R-P* curve and *T-P* phase diagram of 2*H*-$MoS_2$ are shown in Fig. 3a and Fig. 3b, respectively. The resistance undergoes a cascade drop with a negative pressure coefficient. Initially, the resistance diminishes by two orders of magnitude up to 25 GPa indicative of a gradual bandgap collapse in the semiconducting state upon compression at room temperature. Afterwards, the resistance drops by a further order of magnitude up to 60 GPa suggestive of an inhomogeneous emergence of a metallic state within the parent semiconducting state. From 60 GPa up to 90 GPa the resistance levels off as the system becomes predominantly metallic. Beyond 90 GPa the resistance saturates to a constant value implying the presence of a purely metallic state out of which superconductivity emerges at low temperature. Unlike the superconducting dome exhibited by alkali metal intercalated[3] or electrostatically[4] doped 2*H*-$MoS_2$, as well as by Cu-intercalated[11] and physically compressed 1*T*-$TiSe_2$ (ref. 1), our findings point to an extended superconducting phase in heavily compressed 2*H*-$MoS_2$ reminiscent of the *T-P* phase diagram of CDW-bearing 1*T*-$TaS_2$ (refs 2, 12) and of the topological insulator $Bi_2Se_3$ (ref. 13), where $T_c$ remains constant over a large pressure range, suggestive of a constant density of states at the Fermi level. We note that this particular feature is the expected signature for band overlap metallization of a layered compound, where the approximately 2D density of states is independent of carrier density.

Presuming bulk superconductivity in 2$H_a$ phase, the observed $T_c$ is orders of magnitude higher than the BCS value previously estimated via the Allen-Dynes formula with a calculated average electron-phonon coupling parameter $\lambda$ of 0.11 at 100 GPa and 0.33 at 200 GPa, respectively[5]. By applying a reliable first-principles evolutionary search, theoretical calculations[14] predicted a decomposition and phase separation of $2H_a$-$MoS_2$ into MoS and S taking place at 135 GPa, narrowly preempting the structural phase transition to the $P_4/mmm$ phase occurring at 138 GPa. As is well-established, elemental sulfur in this pressure range adopts phase IV and superconducts with a $T_c$ larger than 10 K at pressure above 93 GPa. The calculated electron-phonon coupling parameter $\lambda$ of 0.75 for the $P_4/mmm$ phase gives rise to a superconducting $T_c$ of about 15 K. It is noteworthy that the $T_c$ of both elemental sulfur and the $P_4/mmm$ phase are very close to the observed $T_c$ in our sample. However, this issue is resolved in our complementary high pressure synchrotron x-ray diffraction and theoretical calculations.

To unveil the underlying mechanism behind the pressure-induced superconductivity in $2H$-$MoS_2$, high pressure synchrotron x-ray diffraction was carried out. The diffraction pattern at various pressure at room temperature is shown in Fig. 4a. The structure of $2H$-$MoS_2$ at ambient conditions belongs to the space group $P6_3/mmc$, with Mo atoms and S atoms positioned at 2(b) and 4(f) ($z$=0.625) Wyckoff's positions, respectively. Only the $z$ position of the S atoms is variable in this structure, so the Le Bail method was used for the refinement. The vertical mark in blue, green and red represents the peak position of $2H$-$MoS_2$, Au and W, respectively, in the diffraction pattern at ambient pressure and 155 GPa. A two-phase mixture pattern indicating the occurrence of an isostructural $2H_c$-to-$2H_a$ phase transition below 29 GPa is evident, as denoted by the two arrows of the (002) peak at 29 GPa. This phase transition pressure is consistent with previous theoretical[5] and experimental[8] studies. The bulk modulus of $2H_c$-$MoS_2$ and $2H_a$-$MoS_2$ obtained by fitting the unit cell volume as a function of pressure with third-order Birch-Murnaghan equation of state (EoS) agree well with the previous experimental values[8] and theoretical values in this work, as shown in Fig. 4b. The $2H_a$ structure is confirmed to be quite robust against pressure-induced phase transition, decomposition and amorphization up to 155 GPa, lending compelling proof that the observed superconductivity is the intrinsic quantum state of the $2H_a$ phase.

To further manifest the superconductivity nature of the $2H_a$ phase, theoretical calculations (see Methods for calculation details) were performed to study the structural integrity and electronic band structure of $2H_a$-$MoS_2$ at high pressures. Our calculations correctly reproduce the isostructural $2H_c$-to-$2H_a$ phase transition starting at about 20 GPa (Supplementary Fig. 1), in good agreement with the synchrotron x-ray diffraction data in this work as well as previous theoretical[5] and experimental[8] results. In addition, our calculations predict a $P6_3/mmc$-to-$P_4/mmm$ phase transition at about 138 GPa (Supplementary Fig. 1), consistent with the transition pressure in ref. 13. However, no peak indexing the tetragonal phase is discernable from the diffraction patterns around 140 GPa. As a result, the possibility of superconductivity stemming from the $P_4/mmm$ phase can be excluded. Furthermore, our calculations predict a scenario of decomposition into MoS+S at either 195 GPa or 140 GPa if we presume

MoS$_2$ adopts either the tetragonal *P*$_4$/*mmm* or the hexagonal *P*6$_3$/*mmc* structure above 138 GPa, respectively (Supplementary Fig. 1). However, no experimental diffraction peaks can be indexed to elemental sulfur and cubic MoS, which excludes the possibility of sample decomposition. There might exist a large kinetic barrier against decomposition, rendering the 2*H*$_a$ phase to be kinetically protected. Both calculated pressures inducing decomposition are far above the pressure (~90 GPa) at which superconductivity emerges in our sample. Henceforth, the possibility of observed superconductivity emerging from precipitated elemental sulfur can be ruled out. Finally, our calculations of the electron-phonon coupling parameter *λ* with strong anisotropy (Supplementary Fig. 2) renders a modest superconducting critical temperature far below that observed in the resistance measurements, which calls for more sophisticated calculation methods other than the isotropic one. The pressure-dependent electronic band structure is shown in Supplementary Fig. 3.

Charge-density-waves (CDWs), here of the band overlap kind, also called excitonic insulators[15], can also be suspected to intervene near the bandgap closing in high pressure 2*H*-MoS$_2$, 2*H*-MoSe$_2$ and 2*H*-MoTe$_2$ (refs 5, 6) and may constitute a parent state for superconductivity to set in following the pressure-induced "quantum melting" as shown, for example, in pressure-metallized 1*T*-TiSe$_2$ (ref. 1), 1*T*-TaS$_2$ (ref. 2), and electron doped TiSe$_2$ (ref. 10). Finally, we note that there is no evidence of strong correlations, Mott transitions, and magnetism playing a role in the metallic state.

Our discovery represents an alternative strategy to chemical and electrostatic doping in achieving unprecedented high-$T_c$ superconductivity in pristine CDW-free semiconducting TMDs with an indirect bandgap. Superconductivity with higher $T_c$ in electron donor-intercalated $MX_2$ ($M$=Mo, W; $X$=S, Se) could be expected at much lower pressure as a result of the synergistic role of pressure and chemical doping, both of which facilitate metallization and exceptionally robust superconductivity.

**Methods**

**Crystal growth.** We synthesized $MoS_2$ single crystals using a conventional chemical vapor transport crystal growth method. Stoichiometric amounts of 99.99 % Mo and 99.999 % S (Kojundo Chemical Laboratory Co., Ltd.) powders were mixed and sealed in a quartz ampoule under high vacuum (approximately $10^{-5}$ Torr). The mixture was repeatedly heated and remixed using a muffle furnace at 700 and 1100 °C to make high-quality polycrystalline powders. Bulk single crystals were grown by chemical vapor transport using 99.999% $I_2$ (5 mg/cm$^3$ of ampoule volume) as a transport agent. We did our best to get high-quality crystals by putting excess amount of sulfur to compensate possible sulfur vacancy that induces electron doping, and by performing vacuum annealing after CVT to remove possibly intercalated iodine, the transport agent in the crystal growth.

**High pressure electrical transport measurements.** The experiment was carried out in a screw-pressure-type diamond anvil cell (DAC) made of non-magnetic Cu-Be alloy. The tungsten (W) gasket with initial thickness of 250 μm was preindented by a pair of gem-quality diamond anvils with a top flat of 100 μm in diameter beveled by 8° from a culet of 300 μm in diameter to a pressure of 20 GPa. Afterwards, a hole with a diameter of 200 μm was drilled in the center of the indented W gasket by IR laser ablation. The hole and pit of the indented W gasket was packed with a mixture of epoxy and cubic-BN powder and compressed repeatedly to a pressure of 35 GPa to ensure the compactness and thickness of the insulating gasket. The surface of the W gasket outside the pit was insulated from the electrodes by a layer of Scotch tape. Platinum (Pt) foil with a thickness of 5 μm was cut and tailored into rectangular and triangular shapes to act as current and voltage probes, respectively. The as-tailored electrodes were positioned onto the surface of the insulating gasket in a quasi-four-point-probe array and fixed outside the indentation pit by scotch tape and glued. The electrodes were compressed and sat firmly on the surface of the insulating gasket. A thin sheet sample in rectangular shape cleaved from a bulk single crystal was placed on top of the electrodes. Some ruby powder was sprinkled onto the culet of the upper diamond for pressure calibration. Finally, the measurement was ready to be implemented by squeezing the sample, electrodes and ruby powder between two diamond anvils and clamping the pressure. A direct-current of 1 mA was introduced to the current probes and the voltage drop in the voltage probes was measured to obtain the electrical resistance of the sample. Low temperature measurements were carried out in a custom made cryostat (CRYO Industries of America, Inc.) with base temperature of 1.6 K. For the magnetoresistance measurements, the magnetic field was applied by a superconducting magnet in a cryostat (JANIS Research Company, Inc.). The pressure was determined by the ruby fluorescence scale[16] below 60 GPa and the diamond Raman scale[17] above 60 GPa.

**High pressure synchrotron x-ray diffraction.** The experiment was performed at Sector 16 BM-D, High Pressure Collaborative Access Team (HPCAT) at the Advanced Photon Source (APS) of Argonne National Lab (ANL). A symmetric Mao-Bell diamond anvil cell was used for the experiment. Pressure was generated by a pair of diamond anvils with a 50 μm top flat beveled by 8° from a 300 μm culet. Tungsten gasket was preindented from 110 μm to about 20 μm in thickness and a hole with a diameter of 25 μm was drilled in the center of the preindented area. To be consistent with the resistance measurement condition, no pressure medium was used. Gold particles as pressure marker[18] were loaded with $MoS_2$ powder (Sigma-Aldrich) into the sample chamber. A monochromatic x-ray beam with wavelength of 0.399945 Å was focused onto an area of ~ 5× 5 μm$^2$ on the sample. The diffraction patterns were collected with a MAR 3450 image plate detector. The two-dimensional image plate patterns were integrated into one-dimensional intensity versus 2$\theta$ data using the Fit2D software package[19]. Refinements of the measured X-ray diffraction patterns were performed using the GSAS+EXPGUI software packages[20].

**Theoretical calculations.** We searched for the structures with various stoichiometric $MoS_x$ (x =

1.0, 4/3, 1.5) with simulation cell sizes of 2 formula units (f.u.) at 100 GPa. Our structure searching simulations were performed through the swarm-intelligence based CALYPSO method[21,22] enabling a global minimization of free energy surfaces merging *ab initio* total-energy calculations as implemented in the CALYPSO code. The method is specially designed for global structural minimization unbiased by any known structural information, and has been benchmarked on various known systems.

Total energy calculations and geometrical optimization for these structures were performed in the framework of density functional theory within the Perdew-Burke-Ernzerhof[23] parameterization of generalized gradient approximation as implemented in the VASP (Vienna Ab Initio Simulation Package) code[24]. The projector-augmented wave (PAW)[25] method was adopted with the PAW potentials taken from the VASP library where $4p^65s^14d^5$ and $3s^23p^4$ are treated as valence electrons for Mo and S atoms, respectively. The cutoff energy of 600 eV and appropriate Monkhorst-Pack $k$-meshes were chosen to ensure that all the enthalpy calculations were well converged to less than 1 meV/atom. Van der Waals interactions were included in the vdW-DF moethod[26] in the initial calculations at zero pressure, where they made an important contribution to the interplanar attraction, but were subsequently removed at high pressures.

Electron-phonon coupling (EPC) was calculated using the plane-wave pseudopotential method and density-functional theory within the Perdew-Burke-Ernzerhof parameterization of generalized gradient approximation as implemented in the Quantum Espresso Package (ESPRESSO)[27]. The use of the plane-wave kinetic energy cutoff of 60 Ry and dense $k$-point sampling (12 × 12 × 12), adopted here, was shown to yield excellent convergence. Forces and stresses for the converged structures were optimized and checked to be within the error between the VASP and ESPRESSO code. The Troullier-Martins norm-conserving scheme[28] was used for Mo and S atom. A 6 × 6 × 6 $q$-point mesh in the first Brillouin zone was used in the EPC calculation. A MP grid of 12 × 12 × 12 was used to ensure $k$-point sampling convergence with a Gaussian width of 0.015 Ry, which approximated the zero-width limits in the calculations of the EPC parameter $\lambda$.

**Acknowledgements**

Z.H.C. is supported by the National Natural Science Foundation of China (51372249) and the Director's Grants of Hefei Institutes of Physical Science, Chinese Academy of Sciences (YZJJ201313). F.Y. is supported by the National Natural Science Foundation of China (11374307). Y.J.Z is supported by JSPS through a fellowship for young researchers and the leading graduate schools ALPS. Z.R.Y is supported by the National Key Basic Research of China (2011CBA00111) and the National Natural Science Foundation of China (U1332143). Y.I. is supported by the Grant-in-Aid for Specially Promoted Research (2500003). Y.I. and T.K. are supported by the Strategic International Collaborative Research Program (SICORP, LEMSUPER) from Japan Science and Technology Agency (JST). J.L.Z and Y.S.Z are supported by High Pressure Science and Engineering Center (HiPSEC), UNLV and sponsored in part by the National Nuclear Security Administration under the Stewardship Science Academic Alliances program through DOE Cooperative Agreement #DE-NA0001982. Z.H.C. acknowledges Xiao-Jia Chen and Alexander Goncharov for the permission to use their instruments and discussions. Z.H.C., T.K. and Y.I. acknowledge discussions with Katsuya Shimizu and Erio Tosatti.

**Additional information**

Supplementary information is available in the online version of the paper. Reprints and permissions information is available online at http://www.nature.com/reprints. Correspondence and requests for materials should be addressed to Z.H.C. (zhchi@issp.ac.cn).

**Competing financial interests**

The authors declare no competing financial interests.

**Figure 1|Temperature-dependent resistance at elevated pressure for 2$H$-MoS$_2$. a**, $R$-$T$ curves below 60 GPa. **b**, $R$-$T$ curves above 90GPa. Dotted arrow indicates the direction of increasing pressure.

**Figure 2|Temperature-dependent resistance under applied magnetic field at 200 GPa for 2$H$-MoS$_2$.** $R$-$T$ curves under applied magnetic field up to 6.5 T. Dotted arrow indicates the direction of increasing magnetic field. Inset: $T$-dependent upper critical magnetic field $\mu_0 H_{c2}$. The dashed line in red indicates the Ginzburg-Landau equation. The dashed line in blue represents the WHH formula.

**Figure 3|Resistance-pressure curve and temperature-pressure phase diagram of 2$H$-MoS$_2$. a,** $R$-$P$ curve at room temperature. **b,** $T$-$P$ phase diagram. "Semi" indicates "semiconductor".

**Figure 4|High-pressure synchrotron x-ray diffraction of 2$H$-MoS$_2$. a,** Representative diffraction patterns at elevated pressures at room temperature. **b,** Unit cell volume as a function of pressure for experiment (open data points) and calculation (curves). The data fitted by a 3$^{rd}$ Birch-Murnaghan equation of state gives ambient pressure volume V$_0$=107.43 Å$^3$, bulk modulus B$_0$=68.84 GPa and its first pressure derivative B$_0$'=4.8 for the 2$H_c$ phase, and ambient pressure volume V$_0$=96.21 Å$^3$, bulk modulus B$_0$=137.94 GPa and its first pressure derivative B$_0$'=3.6 for the 2$H_a$ phase, respectively.

**Figure S1|Calculated enthalpy of phase transition. a,** Enthalpy of $2H_a$-$MoS_2$ relative to $2H_c$-$MoS_2$ as a function of pressure. Our calculation shows that the $2H_c$ phase transforms to the $2H_a$ phase at about 20 GPa. **b,** Enthalpy of $P4/mmm$-$MoS_2$ relative to $2H_a$-$MoS_2$ as a function of pressure. Our calculation indicates that the $2H_a$ phase transforms to the $P4/mmm$ phase at about 138 GPa. **c,** Enthalpy of hypothetical decomposed products relative to $2H_a$-$MoS_2$ as a function of pressure. We calculate the reference enthalpy of $MoS_2$ without considering the $P6_3/mmc \rightarrow P4/mmm$ phase transition. Through CALYPSO code searching, MoS, $Mo_2S_3$ and $Mo_3S_4$ is predicted to adopt the CsCl structure, $P4/mmm$ structure and $I4/mmm$ structure at 100 GPa, respectively. The calculated pressure at which $MoS_2$ decomposes into MoS and S is 140 GPa. **d,** Enthalpy of hypothetical decomposed products relative to $2H_a$-$MoS_2$ as a function of pressure. We calculate the reference enthalpy of $MoS_2$ considering the $P6_3/mmc \rightarrow P4/mmm$ phase transition. The calculated pressure at which $MoS_2$ decomposes into MoS and S is 195 GPa.

**Figure S2| Statistical values of λ (ω) for various *q* points.** EPC calculations and application of the Allen-Dynes modified McMillan equation at 150 GPa yield remarkably low critical temperature $T_c$ of 0.2 K for $2H_a$-$MoS_2$. So, we analyzed the statistical values of λ (ω) at different *q* points. The anisotropic value of λ (ω) may account for the discrepancy between theoretical and experimental $T_c$.

**Figure S3| Calculated electronic band structures at elevated pressures. a,** 10 GPa in $2H_c$ phase. **b,** 50 GPa in $2H_a$ phase. **c,** 80 GPa in $2H_a$ phase. **d,** 100 GPa in $2H_a$ phase. **e,** 150 GPa in $2H_a$ phase. **f,** 200 GPa in $2H_a$ phase. The bandgap diminishes

from 1.29 eV at ambient pressure to 0.4 eV at 10 GPa indicative of semiconductivity and to almost zero at 30 GPa suggestive of semimetallicity. Upon further compression, the metallicity is gradually enhanced due to the progressive overlap of valence band and conduction band.

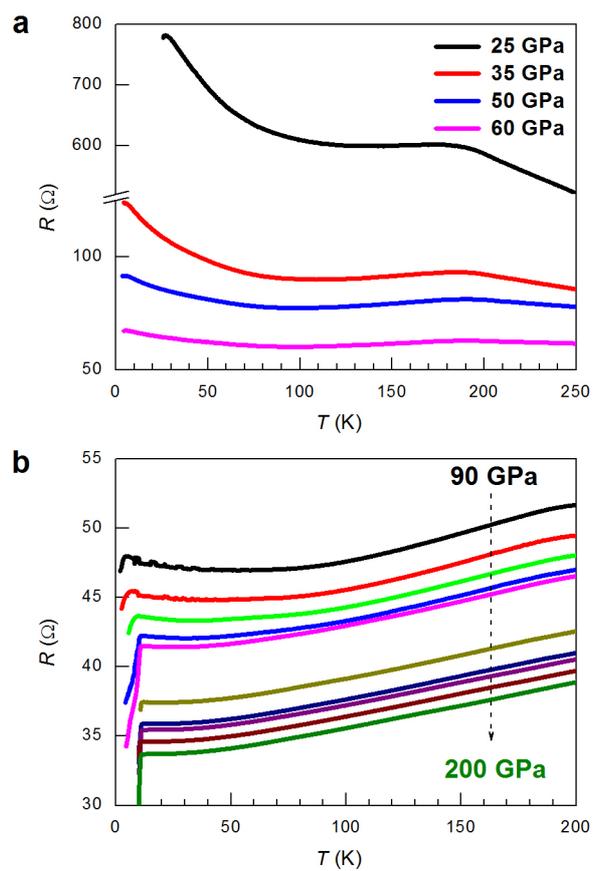

**Fig. 1**

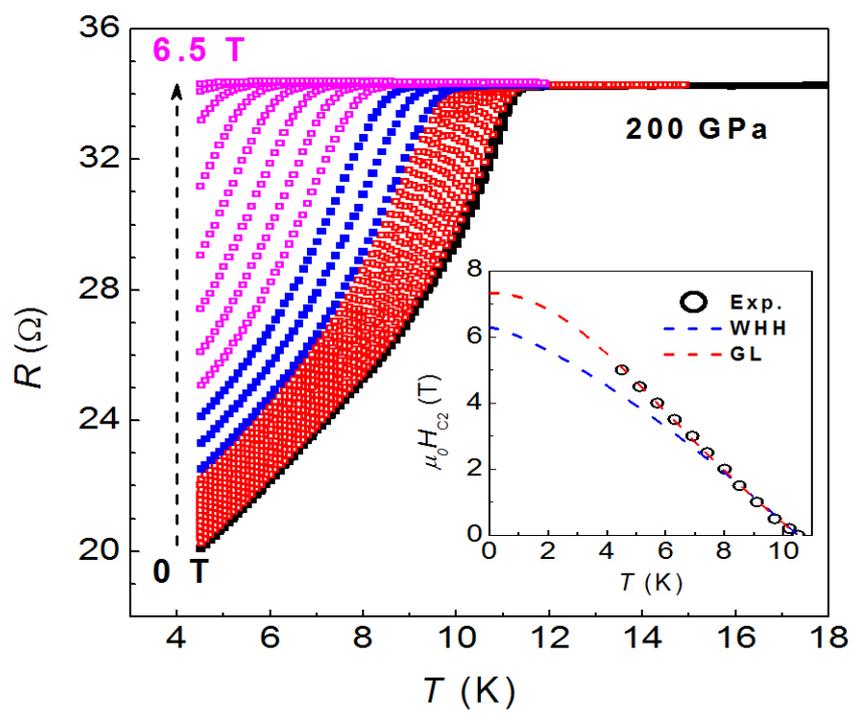

**Fig. 2**

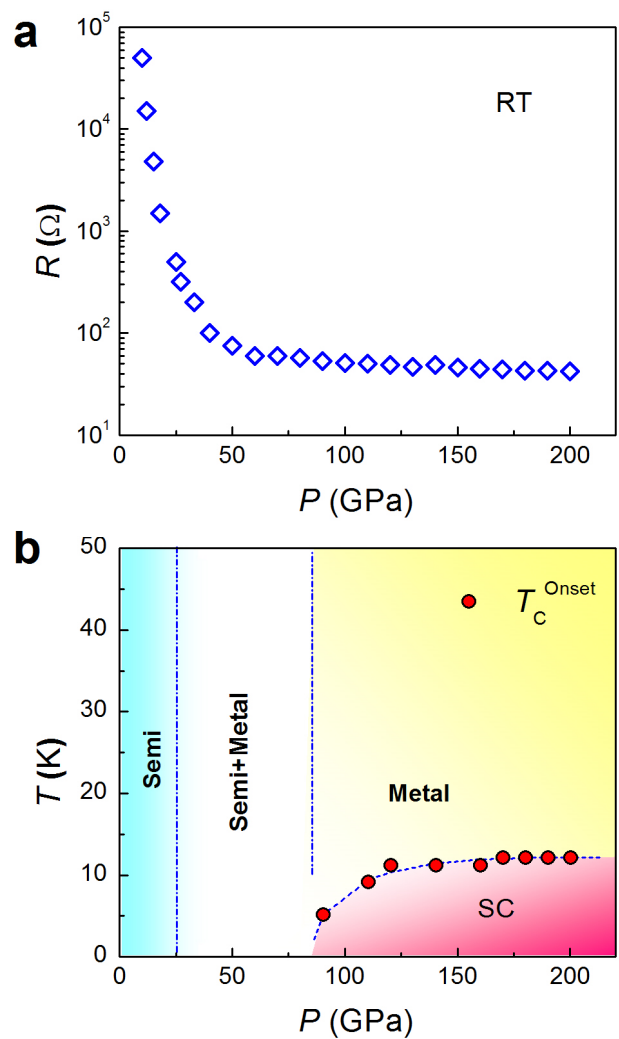

**Fig. 3**

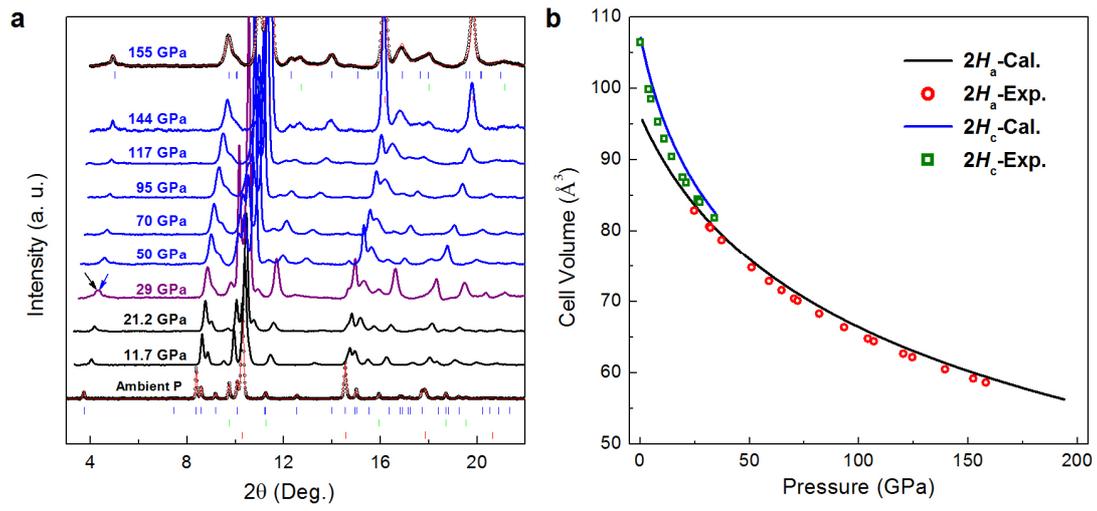

**Fig. 4**

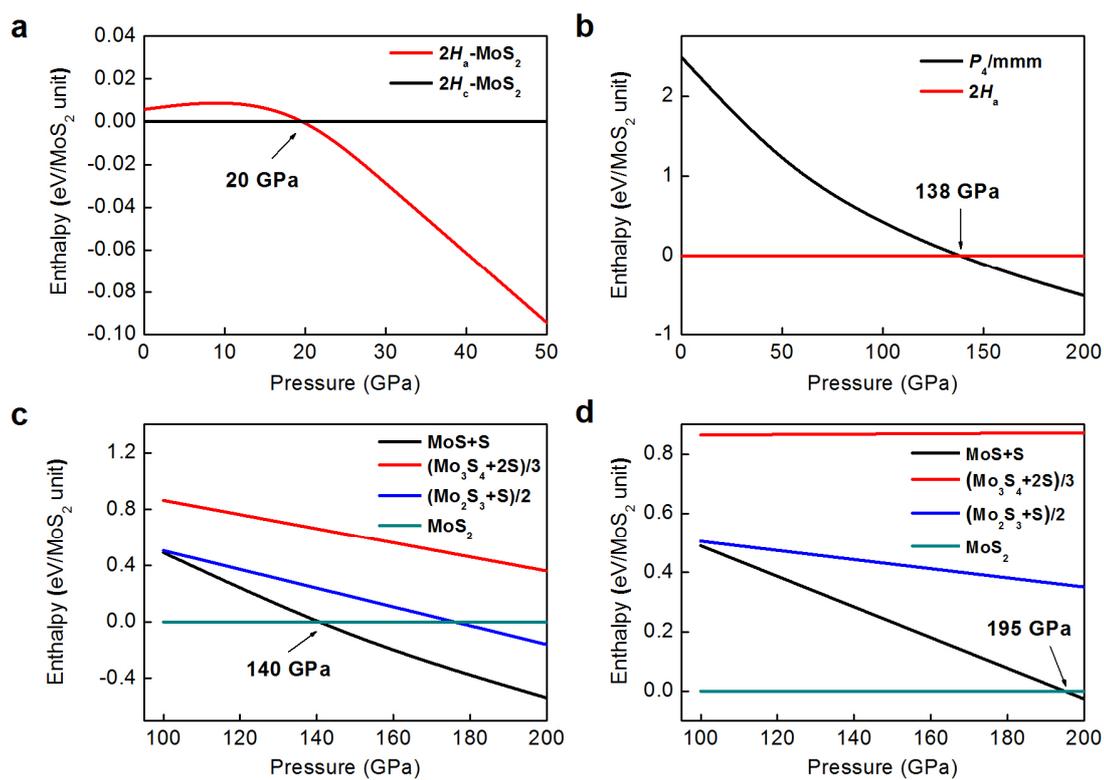

**Fig. S1**

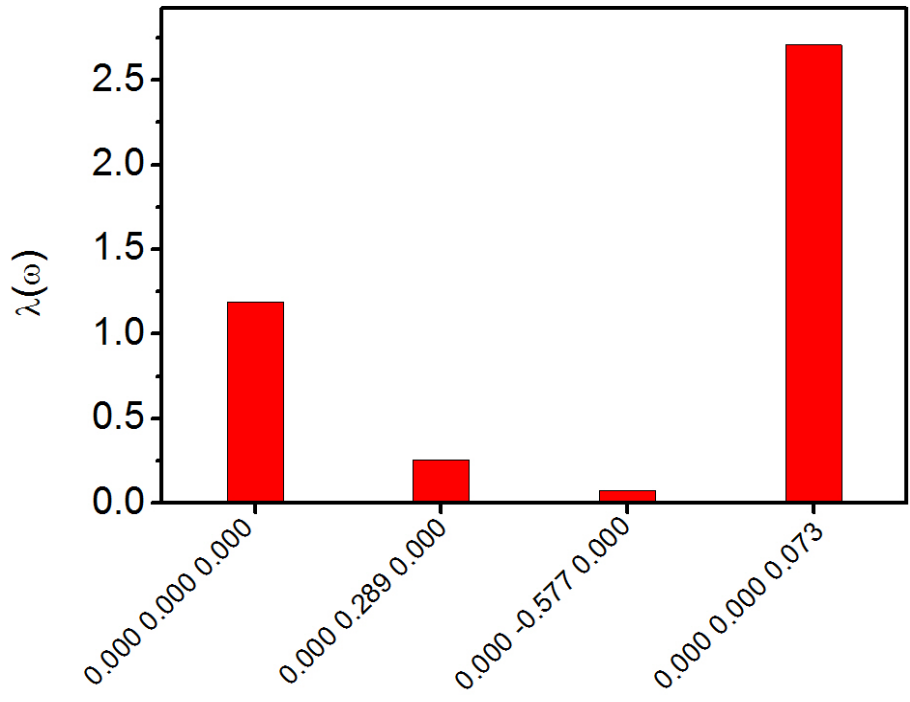

**Fig. S2**

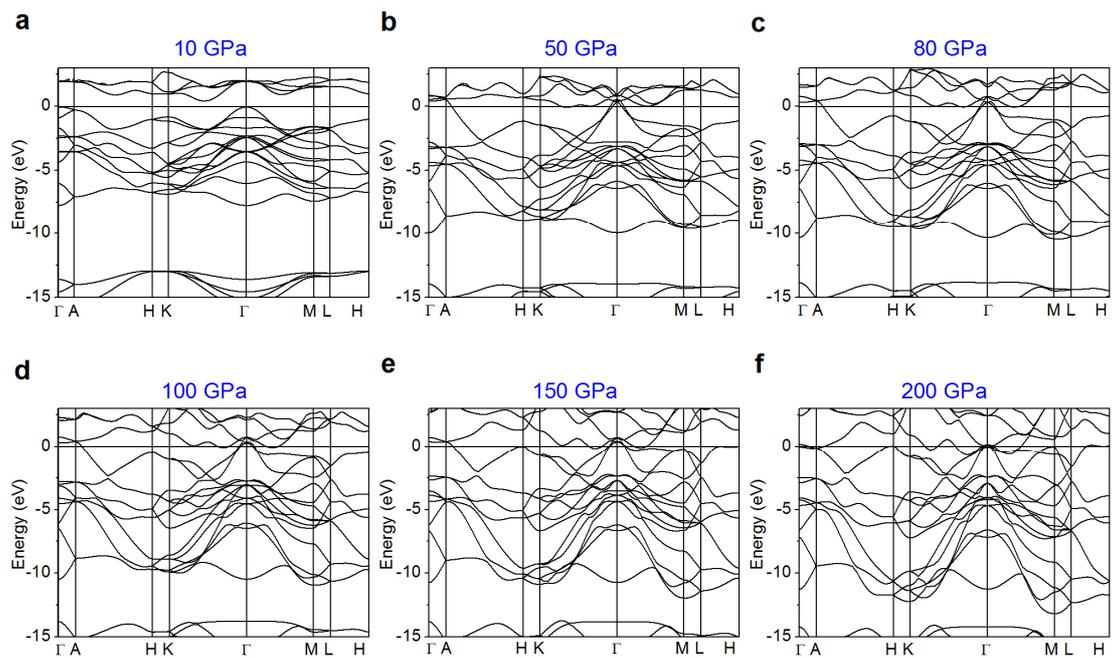

**Fig. S3**